\documentclass[fleqn,11pt]{article}
\usepackage[cp1251]{inputenc}
\usepackage{amsmath,amssymb}

\usepackage{amsfonts,amssymb,cite}
\usepackage{graphicx}

\def\noi{\noindent}

\newcommand{\Title}[1]{\noi {{\Large\bf #1}}\\[1ex]}

\newcommand{\Author}[2]{\noi{\bf #1}\\[2ex]\noi{\normalsize\it #2}\\}

\newcommand{\Abstract}[1]{\vskip 2mm \begin{center}
        \parbox{16.4cm}{\small\noi #1} \end{center}\medskip}
\newcommand{\foom}[1]{\protect\footnotemark[#1]}

\def\nqq{\hspace*{-2em}}

\def\Jl#1#2{#1 {\bf #2},\ }

\def\ApJ#1 {\Jl{Astroph. J.}{#1}}
\def\CQG#1 {\Jl{Class. Quantum Grav.}{#1}}
\def\DAN#1 {\Jl{Dokl. AN SSSR}{#1}}
\def\GC#1 {\Jl{Grav. Cosmol.}{#1}}
\def\GRG#1 {\Jl{Gen. Rel. Grav.}{#1}}
\def\JETF#1 {\Jl{Zh. Eksp. Teor. Fiz.}{#1}}
\def\JETP#1 {\Jl{Sov. Phys. JETP}{#1}}
\def\JHEP#1 {\Jl{JHEP}{#1}}
\def\JMP#1 {\Jl{J. Math. Phys.}{#1}}
\def\NPB#1 {\Jl{Nucl. Phys. B}{#1}}
\def\NP#1 {\Jl{Nucl. Phys.}{#1}}
\def\PLA#1 {\Jl{Phys. Lett. A}{#1}}
\def\PLB#1 {\Jl{Phys. Lett. B}{#1}}
\def\PRD#1 {\Jl{Phys. Rev. D}{#1}}
\def\PRL#1 {\Jl{Phys. Rev. Lett.}{#1}}

\def\lal{&&\nqq {}}

\def\beq{\begin{equation}}
\def\eeq{\end{equation}}
\def\bear{\begin{eqnarray}}
\def\bearr{\begin{eqnarray} \lal}
\def\ear{\end{eqnarray}}
\def\earn{\nonumber \end{eqnarray}}

\begin{document}
\thispagestyle{empty}
\twocolumn[


\Title{Short-wave Approximation For Macroscopic Cosmology with Higgs Scalar Field  \foom 1}

\Author{Yu.G. Ignat'ev and D.Yu. Ignat'ev}
    {Institute of Physics, Kazan Federal University, Kremlyovskaya str., 18, Kazan, 420008, Russia}


\Abstract
 {Based on the macroscopic equations of cosmological evolution obtained earlier by the Author, a closed system of macroscopic
  Einstein equations in the short-wave approximation for perturbations of the scalar Higgs and gravitational fields has been obtained and examined. The obtained exact solutions of macroscopic equations are determined by three microscopic parameters, depending on the spectrum of perturbations.
}
\bigskip

] 

\section*{Introduction}
%
 In \cite{Yu_Pr_19}, a closed system of equations was obtained that determines the macroscopic evolution of the Universe, consisting of locally fluctuating gravitational perturbations and perturbations of the classical scalar field with the Higgs interaction potential, taking into account both transverse and longitudinal field perturbations.
 The obtained macroscopic model consists of a system of ordinary linear homogeneous differential equations describing the evolution of microscopic perturbations of the gravitational and scalar fields on a macroscopic background, and a system of nonlinear ordinary inhomogeneous differential equations determined by macroscopic mean square values of perturbations.
  In the same work, a particular exact solution of this system of equations was obtained for the case of short-wave transverse perturbations, which describes the transition of the macroscopic Universe from the ultrarelativistic stage of expansion to the inflationary one.

The principal complexity of the obtained mathematical model lies in the fact that the macroscopic background relative to which perturbations evolve, according to the method of a self-consistent field, is determined by the same perturbations, therefore, the linear nature of the perturbations is misleading.
 In this regard, it is difficult to uncouple the resulting system of equations. This work is devoted to the derivation of the closed system of Einstein macroscopic equations  based on the formulated model for the space - flat Friedmann Universe in the short-wave approximation for local fluctuations of the scalar and gravitational fields and the solution of these equations.

\section{Macroscopic Model of an Iso\-t\-ro\-pic Space - Flat Universe}
The macroscopic cosmological model formulated in \cite{Yu_Pr_19} includes, first of all, linear equations for perturbations of the Friedmann metric
\begin{eqnarray}
\label{Freed}
ds^2=a^2(\eta)(d\eta^2-dx^2-dy^2-dz^2)
\end{eqnarray}
and scalar field. Further on, we call such equations  \emph{evolution equations of perturbations}. Secondly, the macroscopic cosmological model includes a system of nonlinear equations for the macroscopic background scalar field $\Phi(\eta)$ and macroscopic scale factor $a(\eta)$, which we will call \emph{macroscopic equation of a scalar field} and \emph{macroscopic Einstein equations}.
\subsection{The mathematical model of averaging Einstein's equations}
The general ideology and procedure for obtaining Einstein's macroscopic equations is described in detail in \cite{Yu_GC4_19} and \cite{Yu_Pr_19}. Here we briefly outline the necessary provisions of the averaging model of local plane-wave fluctuations in the case of the classical scalar Higgs field $\Phi $ with the Lagrange function
\begin{equation}\label{Ls}
\mathrm{L}_s=\frac{1}{8\pi}\biggl(\frac{1}{2}g^{ik}\Phi_{,i}\Phi_{,k}-V(\Phi)\biggr),
\end{equation}
where $V(\Phi)$ is the potential energy of the scalar field, which for the Higgs potential takes the form:
\begin{equation}\label{Higgs}
V(\Phi )=-\frac{\alpha }{4} \left(\Phi ^{2} -\frac{m^{2} }{\alpha } \right)^{2} ,
\end{equation}
$\alpha$ is the constant of self-action, $m$ is the mass of scalar bosons.
Moreover, the scalar field equation with respect to the Lagrange function \eqref{Ls} has the form:
\begin{equation}\label{Eq_s}
\Box\Phi+V'_\Phi=0.
\end{equation}
Further
\begin{equation}\label{T_{iks}}
T^i_k=\frac{1}{8\pi}\bigl(\Phi^{,i}\Phi_{,k}-\frac{1}{2}\delta^i_k\Phi_{,j}\Phi^{,j}+\delta^i_k m^2V(\Phi)\bigr)
\end{equation}
there is a tensor of the energy - momentum of a scalar field.

We write the metric with gravitational perturbations with respect to the Friedman metric in the synchronous reference frame (see, for example, \cite{Land_Field}):
\begin{eqnarray}
\label{metric_pert}
ds^2=ds^2_0-a^2(\eta)h_{\alpha\beta}dx^\alpha dx^\beta.;
\end{eqnarray}
Moreover, the covariant perturbations of the metric are equal to:
\begin{equation}\label{dg}
\delta g_{\alpha\beta}=-a^2(\eta)h_{\alpha\beta}.
\end{equation}
Further:
\begin{eqnarray}\label{defh1}
 h^\alpha_\beta=h_{\gamma\beta}g^{\alpha\gamma}_0\equiv-\frac{1}{a^2}h_{\alpha\beta};\\
 \label{defh2}
h\equiv h^\alpha_\alpha\equiv  g^{\alpha\beta}_0h_{\alpha\beta}=
 -\frac{1}{a^2}(h_{11}+h_{22}+h_{33}).
\end{eqnarray}

Each perturbation mode is written in the form:
\begin{equation}\label{h_mode}
h_{\alpha\beta}(\mathbf{r},\eta)=h_{\alpha\beta}(\mathbf{n},\eta)\mathrm{e}^{i\mathbf{nr}+i\psi_0}+\mathbb{CC},
\end{equation}
where $\mathbb{CC}$ means the complex conjugate, and $\psi_0$ is an arbitrary constant phase for each harmonic $\mathbf{n}$.
At each time instant $\eta_0$, a separate independent mode of metric perturbations is completely described by three-dimensional
tensor amplitudes $h_{\alpha\beta}(\mathbf{n},\eta_0)$, the classification of which is in the direction of the spatially similar vector
$\mathbf{n}$ and the three - dimensional Kronecker tensor $\delta_{\alpha\beta}$ is given by Lifshitz\cite{Lifshitz}. The vector, transverse and scalar perturbations correspondingly, in  accordance with the Lifshitz classification; the notation of the perturbations coincides with the standard one \cite{Land_Field}.

For transverse disturbances
\begin{equation}
\label{h(eta)}
h_{\alpha\beta}=e_{\alpha\beta}S\mathrm{e}^{i\mathbf{nr}+i\psi_0}+\mathbb{CC},
\end{equation}
Where $S(\mathbf{n},\eta)$ is the amplitude of the perturbations, so that
\begin{eqnarray}
\label{perp}
h^\alpha_\beta n_\alpha=0;\\
\label{perp1}
h=0.
\end{eqnarray}
For vector perturbations of the metric:
\begin{eqnarray}\label{vect}
h_{\alpha\beta}=V_\alpha n_\beta+V_\beta n_\alpha; \quad (\mathbf{nV})=0;\\
\label{ortov}
\mathbf{V}=\mathbf{v} \mathrm{e}^{i\mathbf{nr}+i\psi_0}+\mathbb{CC},
\end{eqnarray}
where $\mathbf{v}(\mathbf{n},\eta), \mathbf{v}(\mathbf{n},\eta)$ is the amplitude of the vector perturbations.
Similarly for longitudinal perturbations of the metric:
\begin{eqnarray}\label{long}
h_{\alpha\beta}=\mathrm{e}^{i\mathbf{nr}+i\psi_0} \bigl(\lambda P_{\alpha\beta}+\mu Q_{\alpha\beta}\bigr) +\mathbb{CC},
\end{eqnarray}
where $\lambda(\mathbf{n},\eta),\lambda^*(\mathbf{n},\eta),\mu(\mathbf{n},\eta),\mu^*(\mathbf{n},\eta)$ is the amplitude of the scalar perturbations
\begin{equation}
P_{\alpha\beta}=\frac{1}{3}\delta_{\alpha\beta}-\frac{n_\alpha n_\beta}{n^2};\quad Q_{\alpha\beta}=\frac{1}{3}\delta_{\alpha\beta}.
\end{equation}
Further, the potential of the scalar field $\Phi(\eta,\mathbf{r})$ is represented in a similar form:
\begin{equation}\label{Ff}
\Phi(\eta,\mathbf{r})\rightarrow \Phi(\eta)+\phi(\eta)\mathrm{e}^{i\mathbf{nr}+i\psi_0} +\mathbb{CC}.
\end{equation}
Then, the Einstein equations with the $\Lambda$ - term and the scalar field equations are averaged over arbitrary wave vectors $\mathbf{n}$ and random phases $\psi_0$ in the second order of the perturbation theory according to the procedure described in detail in the works cited above \cite{Yu_GC4_19} and \cite{Yu_Pr_19}.

\subsection{Evolution Equations for Perturbations}
The evolution equation for the amplitudes of perturbations of the scalar field  $\phi(\eta)$ has the form:
\begin{eqnarray}\label{Eq_phi}
\phi''+2\frac{a'}{a}\phi'+\hskip 5cm \nonumber\\
a^2\phi\biggl(m^2+\frac{n^2}{a^2}-3\alpha\Phi^2\biggr)+\frac{\mu'}{2}\Phi'=0,
\end{eqnarray}
where $\alpha$ is the self-action constant in the Higgs potential, $m$ is the mass of the scalar field, $\mathbf{n}$ -- is the perturbation wave vector.
Let us write out  the following nontrivial combinations of Einstein equations as linearly independent evolution equations for the amplitudes of perturbations of the gravitational field $v(\eta)$, $S(\eta)$, $\lambda(\eta)$, $\mu(\eta)$ \footnote{Details see in \cite{Yu_Pr_19}}:

\begin{eqnarray}
\label{24}
v'=0;\\
\label{22-11}
S''+2 \frac{a'}{a}S'+n^2S=0;\\
\label{34}
\frac{1}{3}(\lambda+\mu)'+\phi\Phi'=0;\\
\label{2*33-11-22}
\lambda''+2\frac{a'}{a}\lambda' -\frac{n^2}{3}(\lambda+\mu)=0;\\
\label{11+22+33}
\mu'' +2\frac{a'}{a}\mu'+\frac{n^2}{3}(\lambda+\mu)\nonumber\\
 +3\phi'\Phi'-3a^2\Phi\phi(m^2-\alpha\Phi^2)=0.
\end{eqnarray}
In this case the equations \eqref{24} and
\eqref{22-11} for the amplitudes of vector and tensor perturbations are independent from other perturbations and coincide with the corresponding equations of the Lifshitz theory. The equation \eqref{2*33-11-22} coincides with the corresponding equation of the standard perturbation theory (see \cite{Land_Field}), while the equation which is identical to  \eqref{34} in the standard perturbation theory determines the longitudinal component of the fluid velocity. The equation \eqref{11+22+33} is also similar to the corresponding second-order equation of the standard perturbation theory (see \cite{Land_Field}).
\subsection{Macroscopic Scalar Field Equation and Einstein Macroscopic Equations}
In \cite{Yu_Pr_19} the following macroscopic equation of the scalar Higgs field
\begin{eqnarray}\label{EqEinst_Macr_Phi}
\Phi''+2\frac{a'}{a}\Phi' +a^2\Phi(m^2-\alpha\Phi^2)-\nonumber\\
\Phi'\biggl[\overline{(SS^*)'}+\frac{1}{3}\overline{(\lambda\lambda^*)'}
+\frac{1}{6}\overline{(\mu\mu^*)'}\biggr]-6a^2\alpha\overline{\phi\phi^*}\nonumber\\
+\frac{1}{2}\overline{\phi'\mu'\ \!^*+\phi'\ \!^*\mu'}- \frac{1}{2}n^2\overline{(\phi\mu^*+\phi^*\mu)}=0
\end{eqnarray}
and the macroscopic Einstein equation for the component  $^4_4$ were obtained,

\begin{eqnarray}\label{EqEinst_Macr_4}
3\frac{a'^2}{a^4}-\frac{\Phi'^2}{2a^2}-\frac{m^2\Phi^2}{2}+\frac{\alpha\Phi^4}{4} - \Lambda =&\nonumber\\
\frac{1}{a^2}\biggl[n^2\biggl(\frac{\overline{SS^*}}{2}+\frac{\overline{\lambda\mu^*+\lambda^*\mu}}{9}-\frac{\overline{\lambda\lambda^*}}{18}
\biggr)+\frac{\overline{S'S'\ \!^*}}{2}&\nonumber\\
+\frac{\overline{\lambda'\lambda'\ \!^*}}{6}-\frac{\overline{\mu'\mu'\ \!^*}}{6}
+2\frac{a'}{a}\overline{(SS^*)'}+\frac{1}{3}\frac{a'}{a}\overline{(\mu\mu^*)'}\biggr]&\\
+\frac{\overline{\phi' \phi'\ ^*}}{a^2}
+\overline{\phi\phi^*}\biggl(\frac{n^2}{a^2}+m^2-3\alpha\Phi^2\biggr),&\nonumber
\end{eqnarray}
where $\Lambda$ is the cosmological constant. Here the Einstein macroscopic equations for the space components $^\alpha_\alpha$ are differential-algebraic consequences of the equations \eqref{EqEinst_Macr_Phi} and \eqref{EqEinst_Macr_4}. Further, $\overline{FF^*}$ means the operation of averaging over the random phase of the oscillations $\psi$ (see \cite{G&C16}):
\begin{equation}\label{average}
\overline{FF^*}=\frac{1}{2\pi}\int\limits_0^{2\pi}FF^* d\psi.
\end{equation}
If there is not only a single isotropic mode of oscillations with the wave vector $\mathbf{n}$,  but a certain spectrum of oscillations $f(n)\geqslant 0$ is given where
\[\int\limits_0^\infty f(n)dn=1,\]
one should also carry out the averaging over frequencies in the macroscopic equations \eqref{EqEinst_Macr_Phi} -- \eqref{EqEinst_Macr_4} :
\[\int\limits_0^\infty F(n,\eta)f(n)dn.\]

Let us note that the operation of averaging over the directions of the wave vector $\mathbf{n}$ defined in  \cite{Yu_Pr_19} implies that
the factor $n^2$, appearing in the spherical coordinate system is transferred to the spectral function $f(n)$. Therefore, when proceeding to averaging of the
specific perturbations, we should take into account this factor and determine the operation of averaging over the spectrum as follows:
\begin{equation}\label{<FF*>}
\overline{F}\equiv \left\langle{FF*}\right\rangle=\int\limits_0^\infty F(n,\eta)f(n)n^2dn.
\end{equation}

So, further one we determine the macroscopic cosmological model with a help of the evolution equations \eqref{Eq_phi}, \eqref{22-11} -- \eqref{11+22+33} and macroscopic equations \eqref{EqEinst_Macr_Phi} -- \eqref{EqEinst_Macr_4}.

\section{The Solution $\Phi\approx \mathrm{Const}$}
As we noted above, the specifics of the self-consistent field method is that microscopic and macroscopic equations should be solved together. Let us consider one of these possible solutions. We start with examining the equations \eqref{EqEinst_Macr_Phi}, \eqref{Eq_phi}, \eqref{34} and \eqref{11+22+33}. It can be noted that when discarding quadratic fluctuations in the inhomogeneous term of the macroscopic scalar field equation \eqref{EqEinst_Macr_Phi}, a constant nonzero solution of this equation
\begin{equation}\label{Phi=Phi_0}
\Phi=\pm \frac{m}{\sqrt{\alpha}}\equiv \Phi_0, \quad (\alpha>0)
\end{equation}
significantly simplifies the studied system of evolution equations \eqref{Eq_phi}, \eqref{34} -- \eqref{11+22+33}, for which in this case we get:
\begin{eqnarray}
\label{Eq_phi_F0}
\phi''+2\frac{a'}{a}\phi'+\phi\biggl( n^2-2a^2m^2\biggr)=0;\\
\label{(l+m)'=0}
(\lambda+\mu)'=0;\\
\label{(l+m)''}
\lambda''+2\frac{a'}{a}\lambda'-\frac{n^2}{3}(\lambda+\mu)=0;\\
\mu''+2\frac{a'}{a}\mu'+\frac{n^2}{3}(\lambda+\mu)=0.
\end{eqnarray}
Thus, the longitudinal perturbations of the metric and the scalar field in this case are independent, the equation for the transverse waves \eqref{22-11},as before, does not depend on the longitudinal perturbations of the metric and the scalar field. The equations  \eqref{(l+m)'=0} -- \eqref{(l+m)''} are easily integrated:
\begin{equation}
\label{l+m=C}
\lambda+\mu=C_1=\mathrm{Const};
\end{equation}
\begin{eqnarray}
\lambda= C_1\frac{n^2}{3}\int\limits^{\eta} \frac{d\eta'}{a^2(\eta')}\int\limits^{\eta'}a^2(\eta'')d\eta''\nonumber\\
+C_2\int\limits^\eta \frac{d\eta'}{a^2(\eta')}+C_3.\nonumber
\end{eqnarray}
As is known \cite{Land_Field}, solutions of the form \eqref{l+m=C}  can be excluded by admissible transformation of the reference frame. Therefore, in the considered solution, we can put
\begin{equation}\label{l=m=0}
\lambda=\mu=0.
\end{equation}
Therefore, in the approximation being considered  \emph{longitudinal metric perturbations are not generated}.

Substituting further \eqref{Phi=Phi_0} and the obtained solution \eqref{l=m=0} into the macroscopic equation of the scalar field \eqref{EqEinst_Macr_Phi}, we reduce it to the form:
\begin{eqnarray}\label{EqEinst_Macr_Phi_1}
\Phi''+\Phi'\biggl(2\frac{a'}{a}-\overline{(SS^*)'}\biggr) +a^2\Phi(m^2-\alpha\Phi^2)\nonumber\\
=6a^2\alpha\overline{\phi\phi^*}.
\end{eqnarray}
Substituting the solution  \eqref{l=m=0} into the macroscopic Einstein equation \eqref{EqEinst_Macr_4}, we reduce it to the form:
\begin{eqnarray}\label{EqEinst_Macr_4_1}
3\frac{a'^2}{a^4}-\frac{\Phi'^2}{2a^2}-\frac{m^2\Phi^2}{2}+\frac{\alpha\Phi^4}{4} - \Lambda &\nonumber\\
=\frac{1}{a^2}\biggl[\frac{\overline{n^2SS^*}}{2}+\frac{\overline{S'S'\ \!^*}}{2}+&\nonumber\\
2\frac{a'}{a}\overline{(SS^*)'}\biggr]+\frac{\overline{\phi' \phi'\ ^*}}{a^2}
+\overline{\phi\phi^*\biggl(\frac{n^2}{a^2}-2m^2\biggr)}. &
\end{eqnarray}

Let us note that the solution \eqref{Phi=Phi_0} taken as the basis, corresponds to \emph{attracting focus} in the standard cosmological model based on the classical Higgs field \cite{Ign_Kokh}.

\section{WKB - Approximation}
We now consider the WKB approximation in the evolution equations \eqref{22-11} and \eqref{Eq_phi_F0}.
\begin{eqnarray}\label{WKB}
n\gg \frac{a'}{a},\frac{\Phi'}{\Phi};\; S'\gg \frac{a'}{a}S;\; \phi'\gg \frac{a'}{a}\phi,
\end{eqnarray}
representing the solutions of these equations $f(\eta)$ in the form
\begin{equation}\label{Eiconal}
f=\tilde{f}(\eta) \cdot \mathrm{e}^{i\int u(\eta)d\eta+i\psi}; \quad (|u|\gg 1),
\end{equation}
 $\psi=\mathrm{Const}$  is a random phase of oscillations (see \cite{G&C16}), and $\tilde{f}(\eta)$ and $u(\eta)$ are functions, weakly changing with scale factor, such that:
\begin{equation}\label{WKB1}
\frac{a'}{a}\sim \frac{1}{\ell};\; \tilde{f}'\sim \frac{\tilde{f}}{\ell}; \; u'\sim \frac{u}{\ell}.
\end{equation}
Further on,  we do not assume that the relation $n\gtrsim am$ is fair, assuming that the value $a(\eta)m$ can be sufficiently great.

\subsection{Tensor Perturbations}
Let us first consider in detail the WKB approximation for tensor perturbations.
Substituting the amplitude of tensor perturbations in the form \eqref{Eiconal} into the equation \eqref{22-11} we obtain, dividing by the WKB approximations in the following order:
\begin{eqnarray}\label{WKB_S0}
(0)\bigr| & (-u^2+n^2)\tilde(S)=0;\\
\label{WKB_S1}
(1)\bigr| & 2u\tilde{S}'+u'\tilde{S}+2\frac{a'}{a}u\tilde{S}=0.
\end{eqnarray}
Thus, we find in the zero approximation from \eqref{WKB_S0}
\begin{equation}\label{u=n}
u=\pm n.
\end{equation}
Substituting the solution \eqref{u=n} into  \eqref{WKB_S1}, we obtain the equation:
\[\tilde{S}'+\frac{a'}{a}\tilde{S}=0,\]
solving which, we finally obtain for the amplitude of tensor perturbations:
\begin{equation}\label{S_WKB}
S= \frac{1}{a}S^0_+ \mathrm{e}^{in\eta+i\psi}+\frac{1}{a}S^0_- \mathrm{e}^{-in\eta-i\psi},
\end{equation}
where $S^0_\pm$ are constant amplitudes, so that $S^0_+S^0_-=|S^0|^2$. Thus, we find in the WKB approximation:
\begin{eqnarray}
SS^*=\frac{1}{a^2}\bigl[|S^0|^2+2S^0_+S^0_-\cos(2n\eta+2\psi) \bigr]\nonumber\\
 \Rightarrow \overline{SS^*}= \frac{|S^0|^2}{a^2};\\
 S'S'\ \!^* = \frac{n^2}{a^2}\bigl[|S^0|^2-2S^0_+S^0_-\cos(2n\eta+2\psi)]\nonumber\\
 +\frac{a'\ ^2}{a^4}\bigl[|S^0|^2+2S^0_+S^0_-\cos(2n\eta+2\psi)] \Rightarrow \nonumber\\
\overline{S'S'\ \!^*}= \frac{1}{a^2}{\overline{|n^2S^0|^2}}+\frac{a'\ ^2}{a^4}\overline{|S^0|^2}\simeq \frac{1}{a^2}{\overline{|n^2S^0|^2}};\\
\label{SS'*}
 (SS^*)'=-2\frac{a'}{a^2}|S^0|^2 \simeq 0.
\end{eqnarray}

\subsection{Perturbations of the Scalar Field}
Let us proceed to scalar perturbations in our model. We now use the relations  \eqref{WKB} -- \eqref{WKB1}  in the equation \eqref{Eq_phi_F0},  dividing it by the orders of the WKB approximation;
\begin{eqnarray}\label{df_0_WKB}
(0)\bigr|: &\tilde{\phi}(u^2-n^2+2a^2m^2)=0;\\
\label{df_1_WKB}
(1)\bigr|: & 2u(a\tilde{\phi})'+a\tilde{\phi}u'=0;
\end{eqnarray}
Thus, we obtain a \emph{dispersion relation} for scalar perturbations
\begin{equation}\label{u}
u=\pm u_0(n)=\pm\sqrt{n^2-2a^2m^2}\quad (\simeq\pm n).
\end{equation}
Solving the simple differential equation  \eqref{df_1_WKB} and substituting the solution in the \eqref{WKB} formulas, we find the WKB-solution of the equation for scalar field perturbations \eqref{Eq_phi_F0}
\begin{equation}\label{sol_WKB-f}
\varphi=\frac{\phi^0_+}{a\sqrt{u_0(n)}}\mathrm{e}^{iu_0(n)\eta}+\frac{\phi^0_-}{a\sqrt{u_0(n)}}\mathrm{e}^{-iu_0(n)\eta}.
\end{equation}
%

Thus, we find the averages:
\begin{eqnarray}\label{<phiphi>}
\overline{\phi\phi^*}\simeq \frac{1}{a^2}\left\langle \frac{|\phi_0(n)|^2}{n^2-2a^2m^2}\right\rangle,\\
\label{<phiphi,n>}
\overline{\phi\phi^*\biggl(\frac{n^2}{a^2}-2m^2\biggr)}\simeq \frac{1}{a^4}\left\langle|\phi_0(n)|^2\right\rangle,\\
\label{<phi'phi'>}
\overline{\phi'\phi^*\ '}\simeq \frac{1}{a^2}\left\langle \frac{n^2|\phi_0(n)|^2}{n^2-2m^2a^2}\right\rangle,
\end{eqnarray}
where it is
\[|\phi_0(n)|^2=(\phi^0_+)^2+(\phi^0_-)^2\]
and to simplify cumbersome expressions, we replaced the sign of the averaging over the spectrum  $\overline{F}\equiv\langle F\rangle$.

\subsection{Macroscopic Scalar Field Equation in the WKB Approximation}
Substituting the relations \eqref{SS'*} and \eqref{<phiphi>} in the macroscopic field equation \eqref{EqEinst_Macr_Phi_1}, let us reduce it to the explicit form:
\begin{eqnarray}\label{EqEinst_Macr_Phi_2}
\Phi''+2\frac{a'}{a}\Phi' +a^2\Phi(m^2-\alpha\Phi^2)=\sigma,\end{eqnarray}
where it is
\begin{equation}\label{sigma}
\sigma=6\alpha\left\langle \frac{|\phi_0(n)|^2}{n^2-2a^2m^2}\right\rangle.
\end{equation}
Thus, in the WKB approximation, the equation for a macroscopic scalar field becomes an inhomogeneous one, with a scalar source in the right part.
Such a field can be interpreted as a charged scalar field, where the charge density $\sigma$  is determined by the spectral function $|\phi_0(n)|^2$. Let us note that the cosmological models with charged scalar fields were investigated in \cite{Yu_Sash_Mish_Dim}.  In the area of
\begin{equation}\label{n>>am}
n^2\gg 2a^2m^2
\end{equation}
the effective density of the scalar charge in the WKB approximation tends to zero:
\begin{eqnarray}\label{sigma_n}
\sigma\approx \sigma_0\equiv
6\alpha\left\langle \frac{|\phi_0(n)|^2}{n^2}\right\rangle>0,\quad (n\gg am),
\end{eqnarray}
thus, the approximate solution \eqref{Phi=Phi_0} becomes the asymptotically exact one. Not discarding the low charge density in the WKB approximation,
we can find a new \emph{quasi-constant} solution of the macroscopic field equation\eqref{EqEinst_Macr_Phi_2} which takes the form of an algebraic equation of the third order:
 \begin{equation}\label{Phi^3}
 \alpha\Phi^3-m^2\Phi+\frac{\sigma_0}{a^2}=0.
 \end{equation}
The equation\eqref{Phi^3} depending on the size of the dimensionless parameter
 \begin{equation}\label{param}
 \gamma^2=\frac{\sigma_0\sqrt{\alpha}}{a^2m^2}
 \end{equation}
 can have the following real solutions:  one negative at $\gamma^2>4/27$, one negative and one doubly degenerate positive solution at  $\gamma^2=4/27$ and, finally, one negative and two positive solutions at $\gamma^2<4/27$. At $n\gg1$ we get the approximate solutions
 \begin{equation}\label{dPhi}
 \Phi\approx\Phi_0-\frac{\sigma_0}{2m^2a^2},
 \end{equation}
  which, with the growth of the scale factor, asymptotically tend to unperturbed solutions \eqref{Phi=Phi_0}.

 At $1\ll n^2<2a^2m^2$ the effective density of the scalar charge becomes constant and negative::
\begin{eqnarray}\label{EqEinst_Macr_Phi_32}
\Phi''+\frac{a'}{a}\Phi'2 +a^2\Phi(m^2-\alpha\Phi^2)\nonumber\\
=-6\frac{\alpha a^2}{m^2}\left\langle |\phi_0(n)|^2\right\rangle.
\end{eqnarray}
In this case, a strict renormalization of the unperturbed constant solutions becomes possible \eqref{Phi=Phi_0}:
\begin{equation}\label{dPhi0}
\Phi\approx\Phi_0+3\frac{\alpha}{m^4}\left\langle |\phi_0(n)|^2\right\rangle.
\end{equation}

\subsection{Einstein Macroscopic Equation in the WKB Approximation}
Substituting the expressions for the averages \eqref{SS'*},  \eqref{<phiphi,n>} and \eqref{<phi'phi'>} in the equation \eqref{EqEinst_Macr_4_1}, let us reduce it to the form:
\begin{eqnarray}\label{EqEinst_Macr_4_2}
3\frac{a'^2}{a^4}-\frac{\Phi'^2}{2a^2}-\frac{m^2\Phi^2}{2}+\frac{\alpha\Phi^4}{4} - \Lambda
=&\nonumber\\
\frac{\left\langle n^2|S_0(n)|^2\right\rangle}{a^4}+\frac{2}{a^4}\left\langle \frac{n^2-m^2a^2}{n^2-2m^2a^2}|\phi_0(n)|^2 \right\rangle.  &
\end{eqnarray}
In particular, in the \eqref{n>>am} limit we obtain the macroscopic Einstein equation in the WKB approximation:
\begin{eqnarray}\label{EqEinst_Macr_4_3}
3\frac{a'^2}{a^4}-\frac{\Phi'^2}{2a^2}-\frac{m^2\Phi^2}{2}+\frac{\alpha\Phi^4}{4} - \Lambda &\nonumber\\
=\frac{1}{a^4}\biggl[\left\langle n^2|S_0(n)|^2\right\rangle +2\left\langle |\phi_0(n)|^2\right\rangle\biggr].  &
\end{eqnarray}

\section{Solutions of Einstein's Macroscopic Equation}
\subsection{Solution in the WKB Limit}
Using the unperturbed solution \eqref{Phi=Phi_0} in \eqref{EqEinst_Macr_4_3} we simplify this equation
\begin{eqnarray}\label{EqEinst_Macr_4_4}
\!\!3a'^2=\tilde{\Lambda}a^4 +
\left\langle n^2|S_0(n)|^2\right\rangle +2\left\langle |\phi_0(n)|^2\right\rangle,
\end{eqnarray}
where
\begin{eqnarray}\label{L->l}
\tilde{\Lambda}=\Lambda+\frac{m^4}{4\alpha}.
\end{eqnarray}

We proceed in the Einstein equation \eqref{EqEinst_Macr_4_4} to the differentiation with respect to the physical time $t$ and the new non-negative variable $u(t)$:
\[t=\int a(\eta)d\eta;\quad u(t)=a^2(t)\geqslant 0\]
-- as a result, we obtain the equation:
\begin{equation}\label{EqEinst_Macr_4_5}
\dot{u}^2=\delta^2u^2+\xi^2,
\end{equation}
where $\dot{u}=du/dt$ and the following denotations are introduced:
\begin{eqnarray}\label{xi,d}
\delta^2=&\displaystyle\frac{4}{3}\tilde{\Lambda};\hspace{5cm}&\\
\xi^2=&\displaystyle\frac{4}{3}\biggl[\left\langle n^2|S_0(n)|^2\right\rangle+2\left\langle |\phi_0(n)|^2\right\rangle\biggr].&\nonumber
\end{eqnarray}

The equation  \eqref{EqEinst_Macr_4_5} is elementarily integrated
\begin{equation}\label{int=t}
\int\limits_{a_0}^a\frac{du}{\sqrt{\delta^2u^2+\xi^2}}=(t-t_0).
\end{equation}
 The integral \eqref{int=t} converges for any values of $a_0,t_0$ at $\xi\not=0$ , wherefrom is seen that one can always put $a(t_0)\equiv a_0=0$, i.e.,  the solution always contains a singularity at $\xi\not=0$, to the left of which it does not continue. Using the freedom of $t_0$  choice we choose this instant of time for the beginning of the timescale
\begin{equation}\label{t0=0}
a(0)=0,\quad (\xi\not=0).
\end{equation}
Then, reversing the integral \eqref{int=t}, we find a solution of the Einstein equation:
\begin{equation}\label{a(t)}
a(t)=\sqrt{\frac{\xi}{\delta}\sinh \delta t},\quad (\xi\not=0).
\end{equation}
In the absence of local perturbations, that is, at $\xi=0$ we obtain the well-known inflationary solution from \eqref{int=t}
\begin{equation}\label{inflat}
\displaystyle a(t)=a_0 \displaystyle\mathrm{e}^{\frac{\delta t}{2}},
\end{equation}
which has a singularity in the infinite past $a(-\infty)=0$.
\subsection{Next WKB Approximation}
Let us now keep the first non-vanishing term by smallness of $a^2m^2/n^2$ in the right part of the equation \eqref{EqEinst_Macr_4_2}.
As a result, instead of \eqref{EqEinst_Macr_4_5} we get the equation:
\begin{equation}\label{EqEinst_Macr_4_5_1}
\dot{u}^2=\delta^2u^2+\zeta^2 u+\xi^2,
\end{equation}
where it is
\begin{equation}\label{zeta}
\zeta^2=\frac{8}{3}m^2\left\langle\left|\frac{\phi_0(n)}{n}\right|^2\right\rangle.
\end{equation}
Assuming $\xi\not=0,\zeta\not=0$, we find a solution of the equation \eqref{EqEinst_Macr_4_5_1}, which is also singular:
\begin{equation}\label{a2(t)}
a(t)=\frac{1}{\delta}\sqrt{\displaystyle\sinh\frac{\delta t}{2}\biggl(2\delta \xi\cosh\frac{\delta t}{2}+\zeta^2\sinh\frac{\delta t}{2}\biggr)}.
\end{equation}
\subsection{Examination of the Solutions}
 The solution \eqref{a2(t)} at $\zeta\to 0$  continuously transitions into the solution \eqref{a(t)}, so we carry out the further research with respect of this solution. Near the singularity $t\to0$ we obtain the asymptotics of the solution \eqref{a2(t)}
\begin{equation}\label{a(t->0)}
\left.a(t)\right|_{t\to 0}\simeq \frac{1}{2}\sqrt{t(4\xi+\zeta^2 t)},
\end{equation}
-- the solution behaves like a cosmological solution at the ultrarelativistic stage of expansion at $\xi\not=0$ while the solution behaves like $a\sim t$ that is \emph{quintessence} at $\xi=0$.  Let us note that the asymptotic  \eqref{a(t->0)} is an exact solution of the Einstein equation at $\delta=0\Rightarrow \tilde{\Lambda}=0$. At $t\to\infty$ the soltuion \eqref{a2(t)} goes to the inflation type \eqref{inflat}
\begin{equation}\label{inflat1}
\left.a(t)\right|_{t\to\infty}\simeq \frac{\sqrt{2\delta\xi+\zeta^2}}{2\delta}\displaystyle\mathrm{e}^{\frac{\delta t}{2}}.
\end{equation}

For cosmology, two dynamic functions that are observable astronomical variables are important: the Hubble constant $H(t)$, and the invariant cosmological acceleration, $w(t)$:
\begin{equation}\label{H,w}
H(t)=\frac{\dot{a}}{a};\quad w(t)=\frac{a\ddot{a}}{\dot{a}^2}.
\end{equation}
Calculating the Hubble constant with respect to the solution \eqref{a2(t)}, we get:
\begin{equation}\label{H}
H(t)=\frac{\delta(2\xi\delta\cosh(\delta t)+\zeta^2\sinh(\delta t))}{\displaystyle 4\sinh\frac{\delta t}{2}\biggl(2\xi\delta\cosh\frac{\delta t}{2} + \zeta^2\sinh\frac{\delta t}{2}\bigg)}.
\end{equation}
In particular, at $\delta\to0$ we obtain from \eqref{H} formula that is also an asymptotic form near the singularity $t\to0$, --
\begin{equation}\label{Hdo}
\left.H(t)\right|_{\delta\to0}\simeq \frac{2\xi+\zeta^2 t}{t(4\xi+\zeta^2 t)}.
\end{equation}
At $t\to\infty$ we get the asymptotics from \eqref{H}
\begin{equation}\label{H|t->8}
\left.H(t)\right|_{t\to\infty}\simeq \frac{\delta}{2}.
\end{equation}

The expression for the invariant cosmological acceleration is too cumbersome, therefore, we give only its expression for the case $\delta=0$
\begin{equation}\label{wd0}
\left.w(t)\right|_{\delta\to0}\simeq -\frac{4\xi^2}{(2\xi+\zeta^2t)^2} \quad (w(0)=-1).
\end{equation}
and at $t\to \infty$:
\begin{equation}\label{w|t->8}
\left.w(t)\right|_{t\to\infty}\simeq 1.
\end{equation}

Let us notice that it is possible to get rid of one parameter in the equation \eqref{EqEinst_Macr_4_5_1} at $\delta\not\equiv0$, introducing dimensionless time variable
\begin{equation}\label{tau}
\tau=\delta t.
\end{equation}
Assuming further that $\delta\not\equiv0$, we reduce the equation  \eqref{EqEinst_Macr_4_5_1} to the dimensionless form:
\begin{equation}\label{EqEinst_Macr_4_5_21}
\dot{u}^2=u^2+\tilde{\zeta}^2 u+\tilde{\xi}^2,
\end{equation}
where
\begin{equation}\label{tilde_xz}
\tilde{\zeta}=\frac{\zeta}{\delta},\quad \tilde{\xi}=\frac{\xi}{\delta}
\end{equation}
and the time derivative $\tau$ is noted by a dot. The observable cosmological parameters \eqref{H} are transformed as follows: $H(t)\to \delta h(\tau)$; $w(t)\to w(\tau)$, where $h(\tau)=\dot{a}_\tau/a$. To interpret the results, we also introduce an \emph{invariant curvature} $K$ of the Friedmann space
\[K^2=\frac{1}{6}R^{ijkl}R_{ijkl}=H^2(1+w^2).\]
Thus, we obtain for the invariant curvature in the time variable $\tau$:
\begin{equation}\label{K}
K=\delta h\sqrt{1+w^2},
\end{equation}
where
\[\lim\limits_{\tau\to\infty}K(\tau)=\frac{\delta}{\sqrt{2}}.\]

\subsection{Graphic Illustration}
Below are the graphs of the evolution of magnitudes $H(\tau)$ (Fig. \ref{Fig1}), $w(\tau)$ (Fig. \ref{Fig2}) and $K(\tau)$ (Fig. \ref{Fig3}) depending on the parameters of the cosmological model $\mathrm{P}=[\tilde{\xi},\tilde{\zeta}]$. %
\begin{center}
\begin{tabular}[1]{c}
\parbox{8cm}{\refstepcounter{figure}\label{Fig1}
\includegraphics[width=8cm]{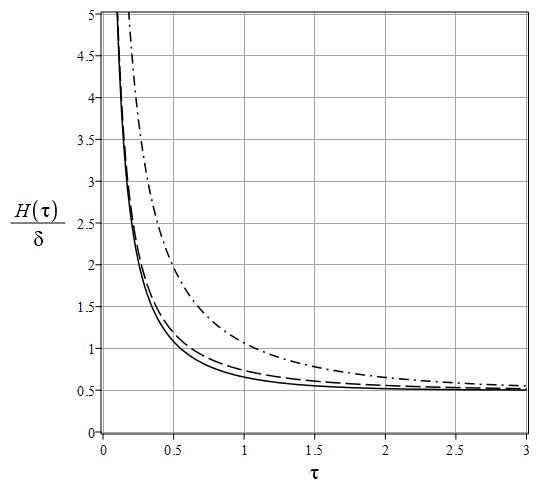}}\\[12pt]
\parbox{8cm}{Figure:\thefigure.\quad Dependence of the behavior of the reduced Hubble constant $H/\delta$ on the model parameters: solid line  -- $\mathrm{P}=[0.001,0.001]$; dotted line -- $\mathrm{P}=[1,1]$; dot-dash line -- $\mathrm{P}=[100,100]$.
} \\
\end{tabular}
\end{center}
\begin{center}
\begin{tabular}[1]{c}
\parbox{8cm}{\refstepcounter{figure}\label{Fig2}
\includegraphics[width=8cm]{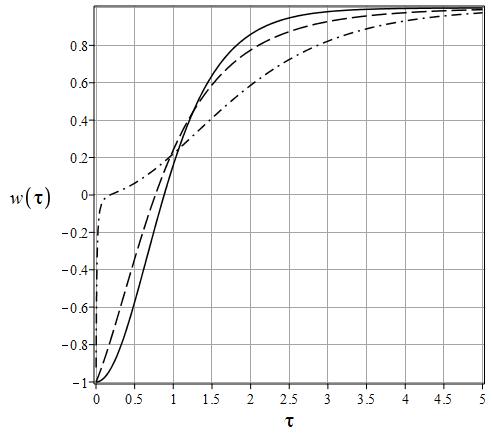}}\\[12pt]
\parbox{8cm}{Figure:\thefigure.\quad Dependence of the behavior of invariant acceleration on the model parameters: solid line -- $\mathrm{P}=[0.001,0.001]$;
dotted line -- $\mathrm{P}=[1,1]$; dot-dash line -- $\mathrm{P}=[100,100]$.
} \\
\end{tabular}
\end{center}
\begin{center}
\begin{tabular}[1]{c}
\parbox{8cm}{\refstepcounter{figure}\label{Fig3}
\includegraphics[width=8cm]{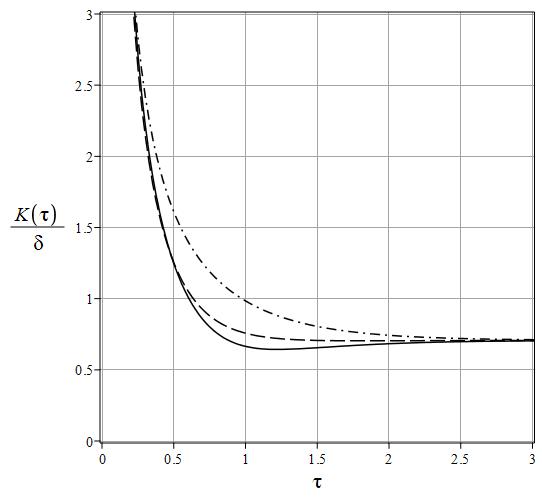}}\\[12pt]
\parbox{8cm}{Figure:\thefigure.\quad Dependence of the behavior of reduced invariant curvature $K/\delta$ on the model parameters: solid line -- $\mathrm{P}=[0.001,0.001]$; dotted line -- $\mathrm{P}=[1,1]$; dot-dash line -- $\mathrm{P}=[100,100]$.
} \\
\end{tabular}
\end{center}
For a correct interpretation of these graphs, it is necessary to take into account the relationship between the variable $\tau$ and the physical time $t$ \eqref{tau} as well as the relations \eqref{tilde_xz} connecting the dimensionless parameters $\delta,\tilde{\xi},\tilde{\zeta}$ with the physical parameters $\tilde{\Lambda},\xi$ \eqref{L->l}, \eqref{xi,d} and $\zeta$ \eqref{zeta}.

\section*{The Discussion of the Results}

When interpreting the results, it is necessary to take into account the fact that the main parameters of the obtained cosmological model $\xi^2$ and $\zeta^2$ should be determined with a help of spectrum averages according to the rule \eqref{<FF*>}. This gives the following refinement of the model parameters:
\begin{eqnarray}\label{<xi^2n^2>}
\xi^2=\frac{4}{3}\int\limits_0^\infty |S_0(n)|^2 n^4 dn+\frac{8}{3}\int\limits_0^\infty |\phi_0(n)|^2 n^2dn;\\
\label{<zeta^2n^2>}
\zeta^2=\frac{8}{3}m^2\int\limits_0^\infty|\phi_0(n)|^2 dn.
\end{eqnarray}
It is also necessary to refine the effective scalar charge density $\sigma$ \eqref{sigma}:
\begin{equation}\label{sigma}
\sigma=6\alpha\int\limits_0^\infty \frac{|\phi_0(n)|^2}{n^2-2a^2m^2}n^2dn.
\end{equation}
In particular, there is the following relation between the model parameters in the range \eqref{n>>am}:
\[\sigma=\frac{9\alpha}{4m^2}\zeta^2\equiv \left(\frac{3\zeta}{2\Phi_0}\right)^2.\]

First, let us note that all the graphs on Fig. \ref{Fig1} -- \ref{Fig3} change their character at $\tau\sim 1$, -- at these values of the time variable, there is a transition from the effective classical braking mode to the acceleration mode with inflationary asymptotic behavior.  Let us remind that the time variable $\tau$  is actually measured in units $1/\sqrt{\tilde{\Lambda}}$, and the value of the effective cosmological constant $\tilde{\Lambda}$ is measured in Planck units according to the accepted in \cite{Yu_Pr_19} system of units. n addition, this value of the cosmological constant refers to the early stages of the evolution of the Universe. If we accept the existing estimates of the radius of curvature of the early Universe at the primary inflation stage $r_k\sim 10\div 10^7 \ell_{pl}$, we obtain an estimate of the value of the primary cosmological constant $\tilde{\Lambda}\sim 10^{-14}\div 10^{-2}\ell^{-2}_{pl}$, where $\ell_{pl}$ is the Planck length. Thus, the change in the acceleration mode can occur in the times $t\sim 10\div 10^7 t_{pl}$. These times are not yet critical from the point of view of observations.

Secondly, let us note that the refinement of the parameters of the macroscopic model $\xi$ \eqref{<xi^2n^2>} and $\zeta$ \eqref{<zeta^2n^2>} can be performed based on the results of quantum theory of generation of scalar and tensor perturbations \cite{starob} (see also \cite{Gorb_Rubak}) which provides the necessary formulas for the spectra of the corresponding perturbations.

Summing up the results of the article, let us note that the article fully implements the program for constructing a macroscopic cosmological model proposed by the Author in the earlier article \cite{Bogolyub}. In this paper, which is the logical conclusion of \cite{Yu_Pr_19}  this program is implemented for the Universe generated by the fluctuating scalar Higgs field. The implementation of this program for the Universe filled with ideal fluid and black holes was carried out earlier in \cite{Yu_GC4_19}.


 \subsection*{Funding}

 This work was funded by the subsidy allocated to Kazan Federal University for the
 state assignment in the sphere of scientific activities.

\newpage

\end{document}